\documentclass[a4paper]{jpconf}
\usepackage{graphicx}
\begin{document}
\title{3D maps of the local interstellar medium: searching for the imprints of past events}

\author{R. Lallement}

\address{GEPI/Observatoire de Paris, CNRS UMR8111, University Paris-Diderot, 5 Pl. J. Janssen, 92195 Meudon, France}

\ead{rosine.lallement@obspm.fr}

\begin{abstract}

Inversion of interstellar (IS) gas or dust absorbing columns measured along the path to stars distributed in distance and direction allows reconstructing the distribution of interstellar matter (ISM) in three dimensions. A low resolution IS dust map based on reddening measurements towards 23,000 nearby stars is used to illustrate the potential of the more detailed maps that are expected within the next several years. The map reveals the location of the main IS cloud complexes up to distances on the order of 600 to 1200 pc depending on directions. Owing to target selection biases towards weakly reddened, brighter stars, the map is especially revealing in terms of regions devoid of IS matter. It traces the {\it Local Bubble} and its neighboring cavities, including a conspicuous, giant, $\geq$1000 pc long cavity in the third quadrant located beyond the so-called $\beta$CMa \textit{tunnel}. This cavity is bordered by the main constituents of the Gould belt, the well-known and still unexplained rotating and expanding ring of clouds and young stars, inclined by $\sim$ 20$^{\circ}$ to the galactic plane. Comparing the dust distribution with X-ray emission maps and IS gas observations shows that the giant cavity contains a large fraction of warm, fully ionized and dust-poor gas in addition to million K, X-ray bright gas. This set of structures must reflect the main events that occurred in the past; today however even the formation of the Gould belt is still a matter of controversy.


 It has been suggested recently that the Cretaceus-Tertiary (KT) mass extinction is potentially due to a gamma-ray burst (GRB) that occurred in the massive globular cluster (GC) 47 Tuc  during its close encounter with the Sun $\sim$70 Myrs ago. Such a hypothesis is based on computations of the cluster and Sun trajectories and the frequency of short GRBs in GC's. Given the mass, speed and size of 47 Tuc, wherever it crossed the Galactic plane it must have produced at the crossing site significant dynamical effects on the disk stars and IS clouds, and triggered star formation. On the other hand, a burst must have produced huge ionization and radiation pressure effects on the ISM. Therefore, identifying (or not) the corresponding imprints should provide additional clues to the extinction source and the ISM history. Interestingly, first-order estimates suggest that the Gould belt dynamics and age could match the expected consequences of the cluster crossing, and that the giant ionized, dust-free cavity could be related to gas ionization and dust evaporation by an intense flux of hard radiation such as produced by a GRB. Moreover, dust-gas decoupling during the crossing and after the burst could have produce a highly inhomogeneous dust to gas ratio, potentially explaining the high variability and pattern of the D/H ratio in the gas phase. Future Gaia data should confirm or dismiss this hypothesis.

\end{abstract}

\begin{figure*}[htbp]
\begin{center}
\includegraphics[width=\linewidth]{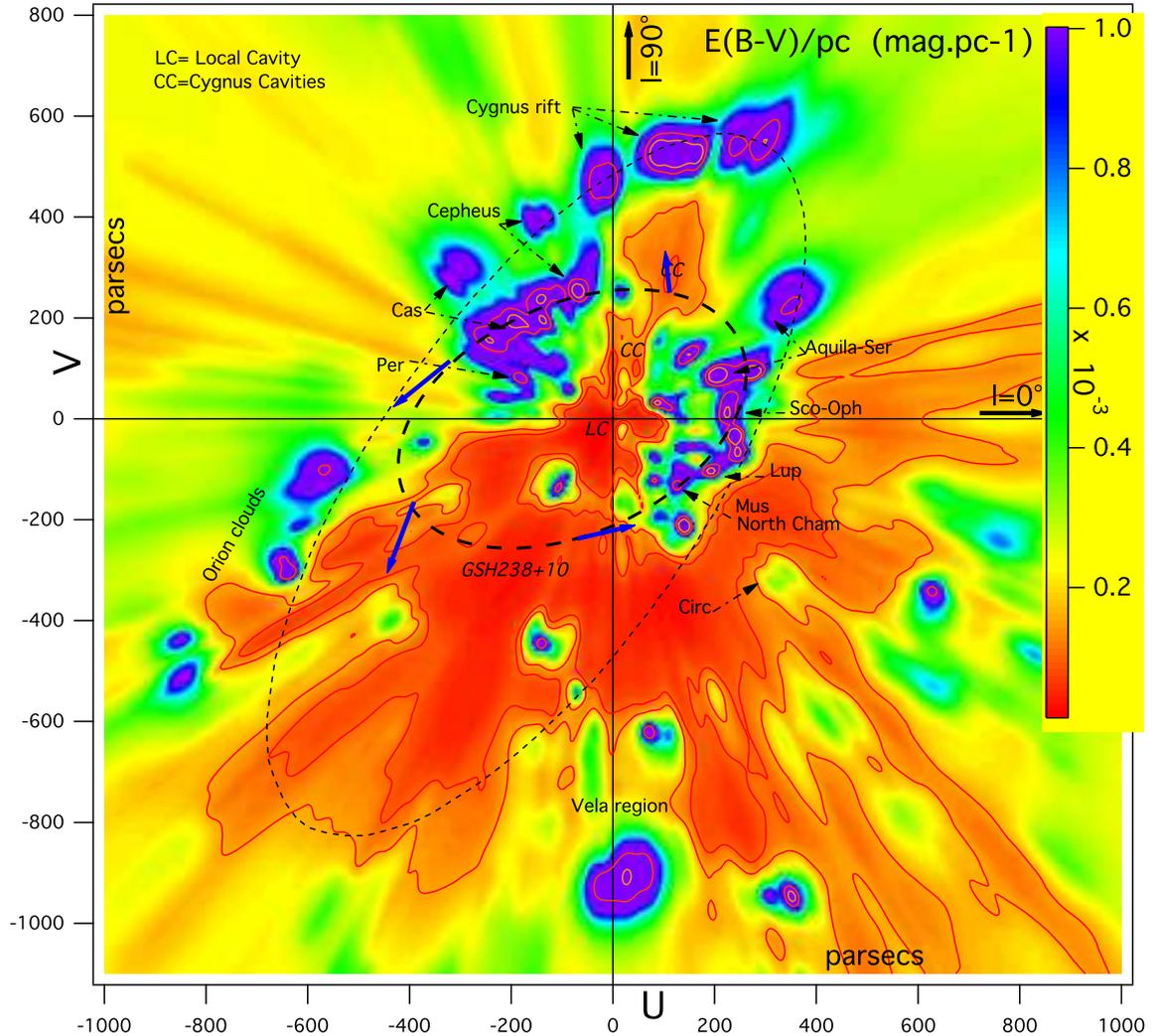}
\caption{Differential color excess in the Galactic plane, derived by inversion of line-of-sight data (adapted from Lallement et al, 2014a). The Sun is at (0,0) and the Galactic center is to the right. Cavities (in red) are probably filled with hot (1 MK) gas or warm, tenuous ionized gas. Also shown is the Gould belt modeled by Perrot and Grenier (smaller ellipse) with four vectors (in blue) illustrating its expansion and rotation. The large ellipse represents the global structure discussed in the text made of the giant cavity in the third quadrant and the Gould belt.}
\label{galplane}
\end{center}
\end{figure*}

\section{3D maps of the local ISM}

Extremely detailed 2D, multi-wavelength observations of the Galactic ISM have been produced, as well as spectral cubes (2D+spectrum) in the radio domain. Still, we lack information on the distance to the observed interstellar structures. The missing information can be obtained by means of inversion of a large series of distance-limited data, i.e. from absorption measurements towards Galactic target stars. Absorption data can be reddening/extinction by dust, gaseous lines, or diffuse bands. The next decade should see significant new information about 3D distributions of gas and dust, owing, first, to the ESA Gaia mission, which will provide precise distances to a billion stars and, second, to the massive photometric/spectroscopic stellar surveys in progress or preparation that benefit from new multiplex techniques and large detectors.  

An example of IS absorption data inversion is shown in Figure 1. A merged catalog of color excess measurements towards 23,000 stars within 2,000 pc has been inverted by means of a Bayesian technique devised by Vergely (2001), to produce a 3D distribution of differential reddening (Lallement et al, 2014). Figure 1 is a horizontal cut in the computed 3D distribution, here along the Galactic plane. Due to the limited dataset, a spatial correlation length must be imposed during the inversion process, resulting in a very poor resolution of the map, on the order of 15 pc, which means that structures of this size of smaller, as e.g. the local and tenuous clouds surrounding the Sun do not appear or are smeared. Instead, what is clearly mapped is the series of well known dense clouds (Scorpius-Ophiucchus, Aquila Rift, Cygnus, Cepheus, etc...) that are indicated in the Figure and are part of the so-called Gould belt (GB). The GB is an elliptical chain of dense clouds and regions of star formation and supernovae (SNRs) production that is contained in a flat disk inclined by  $\sim$ 20$^{\circ}$ to the galactic plane (see e.g. Grenier (2004) for a recent review of the belt properties). The 3D maps show that close to the Plane, in  the first (0-90$^{\circ}$), second (90-180$^{\circ}$) and fourth (270-360/0$^{\circ}$) quadrants the belt is an arc-shaped structure that surrounds the Local Cavity (LC, see below) and two other empty cavities (here called the Cygnus cavities: CC on the map). The Orion and Vela clouds, on the other hand, form extensions of this arc at large distance in the third (180-270$^{\circ}$) quadrant. 
 
The photometric catalogs being limited in brightness, those stars that are weakly reddened also suffer less extinction and are preferentially observed. This results in biases towards low-extinction targets and in turn in a better mapping of the cavities compared to the clouds themselves. Indeed,  the maps show quite well a series of regions devoid of dust or cavities, the $\sim$100 pc wide Local Bubble or Local Cavity (LC) around the Sun (Cox, 1998), and several cavities that are connected to the LC by tunnels including the Cygnus cavities. The Galactic Plane map in Fig 1  also reveals at larger distance in the third quadrant (between galactic longitudes 180 and 270 $^{\circ}$) a conspicuous, very wide region devoid of dust that extends up to at least one  kpc. This huge cavity is connected to the Local Bubble along its most elongated part in the third quadrant, the \textit{region of bizarre emptiness} (Don Cox, 1998), also called the $\beta$CMa tunnel (Welsh, 1991, Gry et al, 1995). A cloud at $\sim$ 200 pc and longitude $\sim$240$^{\circ}$ forms a limited barrier between the two. We identify this huge cavity as the actual 3D counterpart to the giant radio shell GSH238+08+10 observed by Heiles (1998). A number of aspects of this cavity deserve further investigation. First, its axis is the same as that of the arc-shaped Gould belt's most compact fraction, as if  the cavity and the belt were part of an ensemble.  Second, the giant star-forming regions in Orion and Vela are both at the periphery of this super-cavity. Third, it is also aligned with an ionization gradient axis derived from white dwarf observations (shown in Fig 1), an axis directed towards l=$\sim$240$^{\circ}$ (Wolff et al, 1999). Fourth, this cavity must be largely filled with fully ionized warm gas, and only partially by the million-K X-ray emitting gas found in the cavities blown by stellar winds and SNRs, according to the comparisons between ROSAT diffuse X-ray emission maps and the 3D distribution (Puspitarini et al, 2014). Finally, a large quantity of the dust in this cavity has been evaporated, and there is evidence that the dust to gas ratio is generally weak. Figure 2 illustrates some comparisons between IS ionized calcium columns measured  in absorption towards nearby stars and the corresponding columns of dust obtained through integration within the 3D distribution between the Sun and the target star. The third quadrant cavity is characterized by CaII to dust ratios that are much lower than in other regions, with differences that can reach three orders of magnitude. Enrichment in CaII is generally linked to dust evaporation, and high CaII to dust ratios are well known signs of such evaporative processes through the combination of both the dust decrease and the CaII increase. We note that CaII columns for stars in this cavity reach 10$^{12}$-10$^{13}$ cm$^{-2}$, corresponding to 4.5 10$^{19}$-10$^{20}$ H cm$^{-2}$ (using local cloud CaII to H ratios) and to ionized cloud total lengths on the order of 100 to 1000 pc for typical warm gas densities on the order of 0.02-0.05 cm$^{-3}$. This confirms a large filling factor of the ionized gas. Finally, although much less marked, the NaI/dust ratio is also higher in the cavity (Fig 2). Since NaI is predominant in the denser clouds that are embedded in the cavity, this suggests that not only the ionized gas but also the denser, more neutral gas is dust-poor (see Puspitarini et al, 2014).

\begin{figure*}[htbp]
\begin{center}
\includegraphics[width=\linewidth]{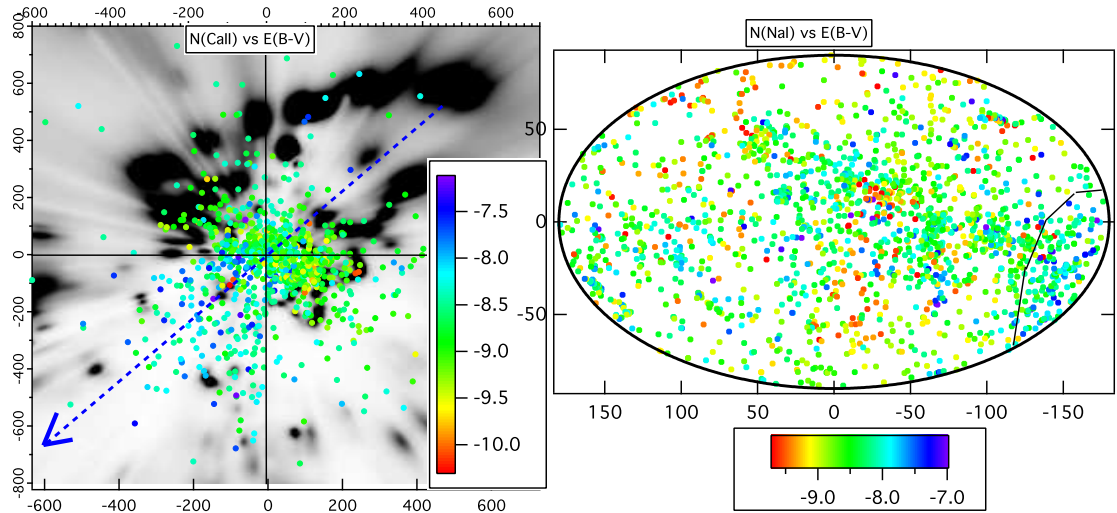}
\caption{Left: Stars close to the Galactic plane for which we have ionized calcium (CaII) absorption measurements (filled circles), superimposed on a subset of the inverted differential color excess distribution (from Puspitarini et al. 2014). The color refers to the ratio between the CaII total column towards the star and the integrated color excess along the same line-of-sight. Stars located in the third quadrant are characterized by a particularly high CaII/E(B-V) ratio, showing that ionized, dust-poor gas is dominant. Right: Full-sky location of neutral sodium target stars. The color refers to the ratio between the NaI column and the integrated color excess. The third quadrant, especially below the GP, is characterized by high NaI/E(B-V) ratios, revealing again dust-poor gas.}
\label{galplane}
\end{center}
\end{figure*}

\section{The Gould belt and the local star properties}

The 3D maps of the multi-phase ISM contain the imprints of its past history, a history that is mainly governed by the star births, lives and deaths by means of their radiation, winds and ejecta, but is also strongly influenced by dynamical events such as waves and tidal effects through their effects on the star formation rate. Hydrodynamical models of the multi-phase gas under the action of winds and supernovae (SNRs) are successful in reproducing most properties and shapes of multi-phase structures such as the Local Cavity, nearby bubbles and clouds (De Avillez \& Breitschwerdt, 2009). Following such models the giant 3rd quadrant cavity that appears in the new maps could be the result of a major series of SNe, and the Vela, Orion and Gould belt structures could represent the next generation of star-forming regions along the compressed and subsequently cooled boundaries of the super-bubble. The large quantity of dust-free, ionized gas may result from the intense radiation field generated by the SNe that formed the giant cavity. 

What is more difficult to reproduce is  the Gould belt itself, which has the distinctive feature of associating a peculiar dynamics of its clouds and young stars with its inclination about the Plane. Over the last 10 years, no consensus has emerged about its origin (and on the contrary some skepticism reappeared about the mere existence of a particular event, see Elias et al, 2009). To recapitulate, two types of scenarios have been modeled and compared to data, the oblique impact of a massive external cloud or a series of explosive events, potentially initiated by a major one, such as a GRB. The former scenario naturally explains the inclination, while the latter is better in that it accounts for the expanding and rotating belt. Developing further the collision scenario, Olano (2001) modeled a 400 pc  wide supercloud (SC) of 2 10$^{7}$ solar masses, initially moving ballistically outside the Plane and slowly rotating, then interacting with a major spiral arm $\sim$100 Myrs ago. During the braking phase the old ($\sim$500 Myrs or more) generation of stars from the SC decoupled from most of the gas, the central part of the SC partly collapsed to form the GB structures while retaining a global rotation, and external parts with lower angular momentum expanded and formed the Local Arm. During the last $\sim$100 Myrs young generations of stars formed and/or exploded  in the compressed GB clouds, forming high pressure cavities and initiating a global expansion of the Belt. Globally, this scenario explains the Sirius supercluster, the Local Arm and the inclined and rotating Gould belt. Of particular interest here, this scenario predicts the formation of a large cavity left behind the braked supercloud, a global pattern that is interestingly  very similar to the 3rd quadrant cavity observed in the prolongation of the Gould belt clouds, as discussed in Lallement et al. (2014a,b). The SC scenario has also been suggested by Lallement et al, (2014) as a source of enhanced D/H variability in the IS gas, following the pioneering work of Linsky et al (2006). If the SC were characterized by a long absence of star formation prior to the collision,  deuterium-rich dust may have gradually formed in it through preferential adsorption of D on dust grains (Draine, 2003). During the braking phase dust-gas decoupling and cloud mixing may have created strong spatial inhomogeneities in the dust to gas ratio, and strong spatial inhomogeneities in the D abundance in the dust grains. Subsequent evaporation of the dust during star formation in OB associations released variable amounts of D and enhanced the D/H variability in the gas phase by comparison with the variability produced by dust evaporation from regions having the same dust to gas ratio and the same initial abundance of D in the grains. This may also explain the observed apparent link between the D/H ratio and the LC/ belt geometry (Linsky et al, 2006).  

Despite all the advantages of the Olano supercloud scenario, there remains one puzzling coincidence, namely that elemental abundances in the super-cloud gas and stars prior to the encounter must be identical to abundances in the disk, given the observed absence of inhomogeneities in chemical abundance in the local gaseous ISM on one hand (e.g. O/H is about constant, see Andre et al (2003), and carbon shows little variability, see Sofia et al, 2004) and the absence of abundance variabilities among the nearby stars (e.g. Nieva et al 2012) whatever their kinematics. The second Gould belt formation scenario, an exploding event (or a series of explosions) with an initial asymmetry has been investigated in detail by Perrot and Grenier (2003) (herafter PG03). By following the evolution of the resulting expanding and rotating ring under the action of the galactic gravitational potential, differential rotation and ISM pressure gradients, PG03 have adjusted the parameters of the initial exploding event to reproduce the locations of the main clouds and OB associations of the belt. After careful examination of a large number of parameters and hypotheses they concluded that is impossible to reconcile the age of the Gould belt inferred from its dynamical structure, on the order of 25 Myrs, with the range of ages of its stars, on the order of 30 to 80 Myrs, in the frame of such models. Moreover, they found that the actual radial velocities of a  number of young structures are in disagreement with the model. 
 
 Recently, the new and massive stellar surveys (SDSS/SEGUE, RAVE, LAMOST/LEGUE) have revealed spatially dependent vertical and radial  motions of the stars within the first 1-2 kpc (see Carlin et al (2013) for a synthetic analysis). The stellar motion pattern is not simple, in particular the articulation with the Gould belt is still unclear. Following previous work by Minchev et al (2009) on high velocity streams and larger timescale, Widrow et al (2014) have devised a model of disk-satellite interaction adapted to the solar neighborhood and conclude that the local disk may be in an unrelaxed mode after the relatively recent crossing of a massive object  with a vertical speed larger than that of the disk stars.  This work considered as crossing candidates a dwarf galaxy or a dark matter subhalo; however a massive cloud as modeled by Olano (2001) is not excluded. A massive globular cluster, as suggested in the next section, would also produce similar non-negligible effects. 
 
\section{Is the Gould belt due to the encounter with the massive globular cluster 47 Tuc?} 


Massive compact globular clusters (GCs) have been recently proposed as the main sources of short/hard gamma ray bursts (GRBs) associated with the merging of neutron star binaries  (see e.g. Gehrels et al, 2009). Because GRBs within $\sim$ 500pc  of the Sun are among the potential sources of mass extinctions through lethal ozone destruction and/or global cooling due to nitrogen dioxide formation (e.g. Melott and Thomas, 2011), estimates of their statistical rate of occurrence in the Sun's vicinity are important clues to the extinction events. For this purpose, Domainko et al (2013) have reconstructed the orbits of the main globular clusters within the last 550 Myr and their distances to the Sun, based on their 3D-space velocities observed today (Dinescu et al 1999) and the solar trajectory parameters (Dehnen and Binney 1998). Interestingly, their study shows that the GRB rate exhibits three peaks at $\sim$ -70, -180 and -340 Myrs, all three due to encounters with the massive cluster 47 Tuc. The most recent peak indeed corresponds to the time period of the well known KT extinction. The range of distances of closest approach is very large due to uncertainties on the present distance and the proper motion of the cluster but values as small as 500-1000 pc are not excluded. 47 Tuc, presently located at 4.5 kpc from the Sun far below the Galactic plane (galactic coordinates ( l, b) = (306$^{\circ}$, -45$^{\circ}$)), is one of the most massive and compact globular clusters, with an estimated mass of 1-2 million $M_{\odot}$ for a diameter of $\sim$ 40 pc. 

Domainko et al (2013) considered primarily the orbitography of the clusters. However, the crossing of the Galactic disk by a GC as massive as 47 Tuc must have produced dynamical effects on the disk stars and gas, and the identification of such effects may reinforce the encounter/GRB scenario (but note that alternatively a close encounter with a GC may have occurred without any associated burst). Two types of question arise: first, does this encounter potentially explain the ordered stellar motions observed today within 1-2 kpc, the Gould belt star and gas properties and the enhanced star formation? And, second, if indeed a GRB occurred within 47 Tuc at a time it was close enough to the Sun, did it influence the formation of the cavities, ionization properties and structures that are seen today in the 3D maps? The answer to those questions evidently requires extensive modeling and is far beyond the scope of this article, however this problem illustrates well the potential use of three-dimensional maps. In what follows we very briefly discuss some general aspects of the encounter and burst.

The first and simplest one, is related to the energy scales. The stellar surface density being on the order of 50 sol.mass pc-2 in the Sun vicinity (Holmberg \&Flynn 2004), a GC mass of 1.5 sol. mass is equivalent to the mass of a cylindric portion of the disk with a 200 pc diameter. According to Dinescu et al (1999) and Domainko et al (2014) the relative velocity between the GC and the Sun for the orbits of close approach is on the order of 100 km.s$^{-1}$, with a vertical component on the order of 50 km.s$^{-1}$.  Since the average vertical velocity of the stars is on the order of 15-20 km.s$^{-1}$, comparing momenta shows that a crossing will generate large perturbations, in agreement with several models of encounters with dwarf galaxies or dark matter subhalos (e.g. Widrow et al, 2014). The width of the impacted area is also far from negligible. For the above cluster speed and mass, the impact parameter that leads to strong (90$^{\circ}$) scattering of a star close to the mid-plane is on the order of 600 pc, according to the free particle approximation formula (2) of Widrow et al (2014). This provides an order of magnitude estimate of the size of the region that is significantly impacted dynamically, an estimated size that is in good agreement with the Gould belt size. A posteriori, it is verified that the diameter of the cluster is much smaller than the spatial scale of the impacted area, justifying models neglecting its finite size. A third aspect is the timescale. The Gould belt stars comprise not only the brightest young ($\leq$ 20 Myr)  massive objects of the main associations, but also stars up to $\sim$ 60-80 Myr old, as shown by Torra et al (2000), Guillout et al (1998) and from open cluster analyses (Bobylev 2006).  If enhanced stellar formation associated with the belt has resulted from strong dynamical perturbations, then this range of GB star ages is in good agreement with a triggering event $\sim$ 70 Myr ago. 
A fourth aspect is the geometry of the crossing. 47 Tuc has a large velocity component in the direction perpendicular to the Plane (Dinescu et al 1999) that we roughly estimate on the order of 50 km.s$^{-1}$. Widrow et al (2014) have studied in detail the influence of the vertical velocity of a crossing object and the subsequent oscillations and show that for the solar neighbourhood such a velocity corresponds to the transition between a simple bending-mode and higher-order modes. This leaves room for the formation of a pronounced \textit{warp} and may explain a structure similar to the Gould plane. The persistence of a significant inclination for the stellar groups is naturally explained by the oscillation time around the disk, on the order of the time elapsed since the crossing, implying that no strong dissipation has occured yet. In the case of the gas, Perrot and Grenier also find an oscillation time on the order of 50 Myr. 

A GRB produces very strong ionization of the surrounding ISM. At variance with long GRBs that occur in dense molecular clouds and ionize large amounts of their surrounding dense gas (see e.g. Krongold \& Prochaska, 2013), the radiation from the short GRBs is entirely available for ionization of gas that is much farther away. With a number of ionizing photons on the order of up to 10$^{58}$  (here we assume that the total radiation of a short GRB is a factor of 1000 less intense than the one of a long GRB and use long GRB numbers from Krongold \& Prochaska, 2013), the burst is able to quasi-instantaneously ionize column-densities on the order of 5 10$^{20}$ cm$^{-2}$ in two cones of less than 50$^{\circ}$ opening angle each. Such columns are similar to the columns of ionized gas in the large 3rd quadrant cavity as discussed above. The effect on the dust must be quite strong too. Such a GRB in mid- or low- density environments must be very effective in destroying the dust, especially the smaller grains, in a large region of space. Crude estimates show that 0.1 micron grains will be evaporated over hundreds of pc where the radiation is not fully attenuated by the interaction with the gas. It is therefore tempting to associate the third quadrant ionized and dust-poor cavity with such a radiation burst. In this case, the ionization gradient (arrow in Fig 2) points to an ionizing source that has most impacted the ISM that is today within the 3rd quadrant cavity. On the other hand, the Sun and the ISM have moved independently since the burst and extrapolating back both motions would require specific models. A crude approximation consists in extrapolating back the Sun location using its present motion in the Local Standard of Rest (LSR), assuming the constancy of this motion over the last 65 Myr, and that the ISM during this period has always moved along with the LSR.  About the second assumption, it is verified today, according to a synthetic study of absorption line Doppler shifts in the spectra of stars in all directions and within $\sim$500 pc (Lallement et al, in prep.), showing that the global motion of the Sun with respect to the IS gas is very similar in direction and modulus to the motion of the Sun in the LSR. Using for this motion the most recent value of Sch{\"o}nrich et al. (2010) puts the Sun at a distance of $\sim$ 1200 pc and coordinates (l, b)= (228$^{\circ}$, -23$^{\circ}$) w.r.t. the present coordinate system. This location corresponds to the extremity of the giant, ionized, dust-free cavity in the third quadrant. Is this is a coincidence or is there a relation of causality between a GRB and the formation of the giant cavity? Evidently more dynamical data and more realistic models are required to answer with confidence. It should be mentioned, finally, that the close encounter with 47 Tuc may have destabilized Oort cloud objects and triggered the infall of comets and meteorites (Montmerle 2014, private comm.) and that those events may have produced the KT extinction without the need for any GRB.

\section{Summary}
Detailed three-dimensional maps of the ISM are expected to be built within the next several years thanks to the massive stellar spectroscopic/photometric surveys in progress or preparation and thanks to the future stellar parallax distance determinations by the ESA mission Gaia (Perryman et al. 2001, de Bruijne 2012). As a precursor of such 3D maps I have presented a recently computed distribution of the differential reddening by the IS dust within $\sim$ 1,000 pc from the Sun. The mapping is based on 23,000 color excess measurements and assigns a location in 3D space to the main dense clouds in the sun vicinity, in particular the structures associated with the Gould belt. The map reveals particularly well the cavities devoid of dust, in particular a giant cavity beyond the so-called $\beta$CMa tunnel, bordered by the Orion and Vela star forming regions. The comparison with IS absorption data show that this cavity contains a large fraction of fully ionized, dust-poor gas. 

Among their many potential uses, such 3D maps may help to study the past history of the nearby ISM.  I discussed an example of a search for the imprints of past events, following the recent suggestion that a close encounter between the Sun and the massive cluster 47 Tuc may have occurred 70 Myr  ago, and that a gamma-ray burst happened at that time in the cluster, potentially at the origin of the KT mass extinction. Interestingly, first-order estimates suggest that the inclined Gould belt of young stars and gas may indeed be related to the dynamical impact of such a crossing event, and that the ionization and dust depletion pattern seen today may have been produced by a short GRB. 

By means of an unprecedented set of  astrometric, spectrometric, and photometric observations  of Milky Way stars Gaia is expected to improve the knowledge of the MW dynamics and evolution. Among all the expected results is the increased accuracy of globular cluster orbital elements and of the local arm motion, allowing one to confirm or dismiss an encounter with a massive globular cluster such as 47 Tuc. Accurate star distances and color excess measurements should allow building much more accurate and detailed 3D distributions of the IS matter, allowing, in turn, identifying in more detail the potential effects of a GC encounter and those of a GRB.\\

Acknowledgements: This work has been partly carried out with support from the French Research Agency (ANR) through the STILISM project. I am indebted to B.R. Sandel who has been an efficient language editor.\\

References:

\end{document}